\documentclass[aps,prl,reprint,superscriptaddress]{revtex4-1}

\usepackage{amsmath}
\usepackage{amssymb}
\usepackage{amstext}
\usepackage{bm}
\usepackage{graphicx}
\usepackage{multirow}
\usepackage{mathtools}

\begin{document}

\title{How Perfect are Perfect Vortex Beams?}
\author{Jonathan Pinnell}
\affiliation{School of Physics, University of the Witwatersrand, Johannesburg 2000, South Africa}\author{Valeria Rodr\'{i}guez-Fajardo}
\affiliation{School of Physics, University of the Witwatersrand, Johannesburg 2000, South Africa}
\author{Andrew Forbes}
\affiliation{School of Physics, University of the Witwatersrand, Johannesburg 2000, South Africa}
\affiliation{Corresponding author: andrew.forbes@wits.ac.za}




\begin{abstract}
Perfect (optical) vortex (PV) beams are fields which are mooted to be independent of the orbital angular momentum (OAM) they carry. To date, the best experimental approximation of these modes is obtained from passing Bessel-Gaussian beams through a Fourier lens. However, the OAM-dependent width of these quasi-PVs is not precisely known and is often understated. We address this here by deriving and experimentally confirming an explicit analytic expression for the second moment width of quasi-PVs. We show that the width scales in proportion to $\sqrt{\ell}$ in the best case, the same as most ``regular" vortex modes albeit with a much smaller proportionality constant. Our work will be of interest to the large community who seek to use such structured light fields in various applications, including optical trapping, tweezing and communications.
\end{abstract}

\maketitle

Vortex modes are structured light fields with a helical (azimuthal) phase of the form $\exp(i\ell\phi)$ that carry $\ell\hbar$ orbital angular momentum (OAM) per photon \cite{Allen1992} and have found many applications \cite{OAM1}. Typically, these modes have a characteristic on-axis intensity null surrounded by a ring of light whose radius scales with the OAM content. Due to the finite numerical aperture of any real optical system, an experimenter is thus limited to a range of usable OAM values, which is of the order of $|\ell| \approx 10^2$ \cite{Roux2003}. Additionally, in certain applications such as optical trapping, it is desirable to have an OAM-containing beam whose width is independent of the OAM content. Perfect Vortex beams (PVs) were introduced as a potential solution to these problems \cite{Ostrovsky13}. Indeed, an ideal PV is an annular ring that can have an arbitrary helical phase whilst maintaining a fixed radius:
\begin{equation} \label{eq:PVideal}
    PV_{\text{ideal}}(r,\phi) \propto \delta(r - R) \, \exp(i\ell\phi)\,,
\end{equation}
where $(r,\phi)$ are transverse cylindrical coordinates and $R$ is the annular ring radius. Such beams are known to be the Fourier transform of the well-studied Bessel beams,
\begin{equation} \label{eq:BB}
    BB(r,\phi) \propto J_\ell(k_r r) \, \exp(i\ell\phi) \,,
\end{equation}
where $k_r$ is the radial wavenumber. An ideal PV, which is dual to an ideal Bessel beam, is an annular ring of infinitesimal thickness of the form of Eq.~\ref{eq:PVideal} with radius $R = k_r f/k$, where $f$ is the focal length of the Fourier lens and $k$ is the wave number of the light. Note that the ring radius is independent of $\ell$ and hence the amount of OAM the beam carries does not influence the beam width. However, an ideal Bessel beam cannot be experimentally realised and so we turn to it's finite-energy approximation: the Bessel-Gaussian (BG) beam \cite{Gori1987} whose field is similar to that of Eq.~\ref{eq:BB} but has an additional Gaussian factor $\exp(-r^2/w_0^2)$, where $w_0$ is the Gaussian width. The PV which is dual to a BG is no longer infinitesimally thin, but instead has thickness $T = 2 f / k w_0$. The complex amplitude of this experimentally realisable quasi-PV is given by \cite{vaity2015perfect},
\begin{equation} \label{eq:PV}
PV(r,\phi) \propto  \exp\left( -\frac{r^2 + R^2}{T^2} \right) \, I_\ell \left( \frac{2 R r}{T^2} \right) \, \exp(i\ell\phi) \,,
\end{equation}
where $I_\ell(\cdot)$ is the modified Bessel function. If the ring radius is much larger than the ring thickness $R \gg T$, or equivalently $k_r w_0 \gg 1$ (in BG parameters), then using the asymptotic form of $I_\ell(\cdot)$, Eq.~\ref{eq:PV} simplifies to,
\begin{equation} \label{eq:PV2}
    PV(r,\phi) \sim \exp\left( -\frac{(r-R)^2}{T^2} \right) \exp(i\ell\phi) \,.
\end{equation}
The above is often quoted as being \textit{the} form of a PV, since the radial amplitude is $\ell$-independent. However, it is only an approximation and so when we refer to a PV we will refer to the true field as given in Eq.~\ref{eq:PV}.

The ring radius of the PV as given in Eq.~\ref{eq:PV} is not OAM-independent, although this is not strikingly apparent when looking at the field structure. A qualitative explanation is that the slope of $I_\ell(\cdot)$ decreases with $\ell$, which shifts the radius where the exponentially decreasing Gaussian term and the exponentially increasing Bessel term intersect. It is this aspect which we aim to quantify here. In the literature, many PV experiments consider topological charges in a small enough range where the increase is unnoticeable/negligible (for example \cite{Sabatyan}). Others have given a semi-empirical rule for the increasing radius (for example \cite{vaity2015perfect}) and some have numerically calculated the OAM-dependent width for the PVs they generate (for example \cite{liang2018rotating}). In some cases, the width increase is even attributed to be some systematic error (for example \cite{Garcia2014simple}). Altogether, it seems apparent that there is a need to make the OAM-dependence of the PV width precise. To the best of our knowledge, we quote here for the first time and confirm experimentally an explicit expression for the OAM-dependent width of quasi-PVs by computing the second moment integrals analytically. We show that in certain regimes, the scaling of the PV width with OAM can be comparable to ``regular" vortex beams and care must be taken in selecting appropriate beam parameters to avoid this. With an analytical expression in hand, we also show how to precisely correct for the increasing width (but only up to a point), thus allowing the dynamic generation of an OAM-containing beam of fixed width. Finally, we argue that since any propagating beam with a helical phase has a vortex core whose radius scales with the topological charge $\ell$, this makes a truly OAM-independent field unattainable.

We begin by presenting an explicit analytical expression for the OAM-dependent width of PVs. There are many different definitions for the width of an arbitrary laser beam profile, but arguably the most convenient definition is the second moment width, defined in cylindrical coordinates as,
\begin{align} \label{eq:SecMomWidth}
 w ^2 = 2\, \frac{\int_0^\infty r\,dr \int_0^{2\pi} d\phi\, r^2 \, |U(r,\phi)|^2 }{\int_0^\infty r\,dr \int_0^{2\pi} d\phi\, |U(r,\phi)|^2} \,,
\end{align}
where $U(r,\phi)$ is the transverse electric field of the laser beam profile. Substituting the field of the PV as given in Eq \ref{eq:PV} into the above yields the following expression for the OAM-dependent width,
\begin{align} 
 w^2(\ell) &= T^2(\ell+1) + R^2 \left( 1 + \frac{I_{\ell+1}\left(\frac{R^2}{T^2} \right)}{I_\ell\left(\frac{R^2}{T^2} \right)}   \right) \,. \label{eq:PVwidth}
\end{align}
It should be noted that $R$ and $T$ are constants which define the ring radius and thickness when $\ell=0$. As $\ell$ changes, so too do these ring attributes. One can verify that when $R/T \gg 1$, the above simplifies to \begin{equation} \label{eq:asympt}
    w^2(\ell) \approx T^2(\ell+1) + 2R^2 \,,
\end{equation}
which shows that even the asymptotic PV field as given in Eq.~\ref{eq:PV2} has an OAM-dependent width. In fact, this asymptotic width scales as $\sqrt{\ell}$ which is the same as for Laguerre-Gaussian (LG) beams. However, since $T$ is small compare to $R$, the width change is slight. One can also verify that in the ideal Bessel beam limit ($w_0\rightarrow \infty$ with $k_r$ fixed or vice versa), Eq.~\ref{eq:PVwidth} reduces to 
\begin{equation}
    w^2(\ell) = w^2 = 2 R^2 \,,
\end{equation} 
which is OAM independent, in agreement with the ideal case.

\begin{figure}[t] 
	\centering
	\includegraphics[width=\linewidth]{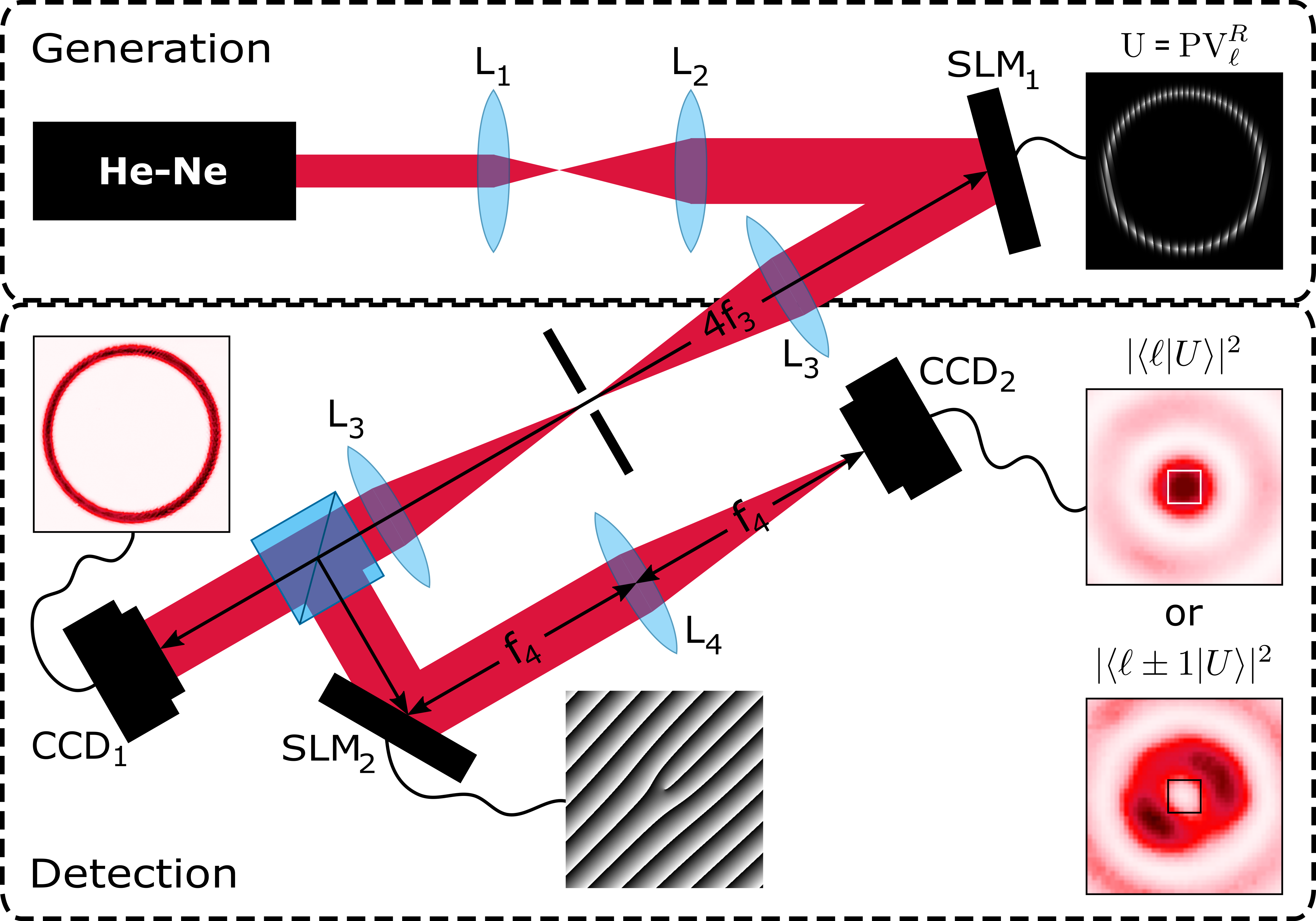}
	\caption{Schematic of the experimental set-up; $L_i$ are lenses of focal length $f_i$, SLM$_i$ are spatial light modulators and CCD$_i$ are CCD cameras. A beam splitter was placed after $L_3$ so that the OAM content of the PV could be verified with SLM$_2$ and CCD$_2$ while the width is calculated from the image taken with CCD$_1$.}
	\label{fig:setup}
\end{figure}

To showcase the OAM-dependent width experimentally, we built the setup shown in Figure \ref{fig:setup}. The beam from a He-Ne laser was expanded and collimated onto a phase-only Holoeye Pluto spatial light modulator (SLM) displaying the necessary digital holograms to generate PVs. Here, we chose to generate the PV field (as given in Eq.~\ref{eq:PV}) directly using complex amplitude modulation \cite{Arrizon2007cam}. The other option is to encode an axicon plus a helical phase onto the SLM (to generate the corresponding BG) and then Fourier transform this field with a lens. Both approaches would yield the same results. Since complex amplitude modulation generates the desired field at the plane of the SLM, it is necessary to relay the field from the SLM to the camera using a 4f lens system. To confirm that the generated PVs had the desired topological charge, we utilised a standard modal decomposition setup \cite{Forbes2016} which is now known to be effective for determining the OAM content of PVs quantitatively \cite{Pinnell:19}.

\begin{figure}[t] 
	\centering
	\includegraphics[width=\linewidth]{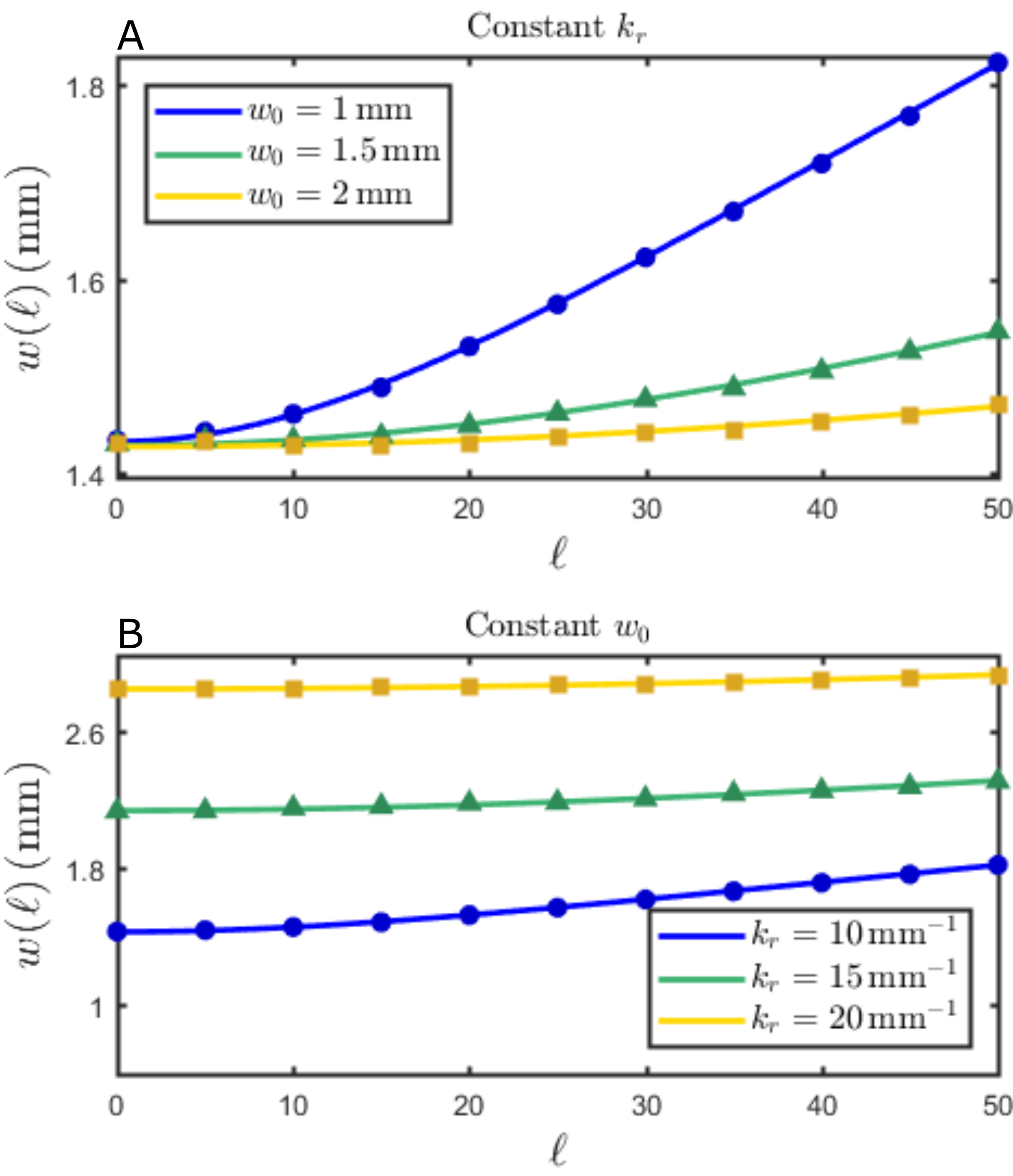}
	\caption{The PV width plotted over $\ell\in[0,50]$ for fixed $k_r = 10 \, \text{mm}^{-1}$ (A) and fixed $w_0 = 1\,\text{mm}$ (B). Lines denote theoretical widths calculated using Eq.~\ref{eq:PVwidth} and symbols denote experimental widths computed from the images taken with CCD$_1$ and using Eq.~\ref{eq:comWidth}. Corresponding curves in A and B have the same value for the product $k_r w_0$ (or equivalently $R/T$) and so have the same width scaling.}
	\label{fig:krw0}
\end{figure}

From the intensity images taken with CCD$_1$ (PointGrey Firefly), the width can be computed numerically in an analogous way to how the second moment integrals are computed, but in Cartesian coordinates,
\begin{align}
w ^2 \approx 4 \frac{\sum_{i=1}^H \sum_{j=1}^V x_i^2 I(x_i,y_j) \Delta x \Delta y }{\sum_{i=1}^H \sum_{j=1}^V  I(x_i,y_j) \Delta x \Delta y}\,, \label{eq:comWidth}
\end{align}
where $H \times V$ are the dimensions of the image in pixels, $(\Delta x,\Delta y)$ are the dimensions of a single pixel and $I(x_i,y_j)$ denotes the intensity value at pixel coordinates $(j,i)$ which are related to the spatial coordinates by
\begin{align}
x_i &= (-x_0 + i)\Delta x  \,, \\
y_j &= (-y_0 + j)\Delta y \,,
\end{align}
and where $(x_0,y_0)$ are the pixel coordinates of the first moment width (essentially the ``centre of mass" of the image). The above numerical approximation of Eq.~\ref{eq:SecMomWidth} is good if the intensity falls to zero sufficiently rapidly at the boundaries of the image and if the pixel dimensions are small compared to the change in intensity between pixels. Care should be taken to remove noise from the image as this will skew the computed width (especially if the noise is far from the origin). Here, this was achieved with background subtraction and median filtering.

\begin{figure}[t] 
	\centering
	\includegraphics[width=\linewidth]{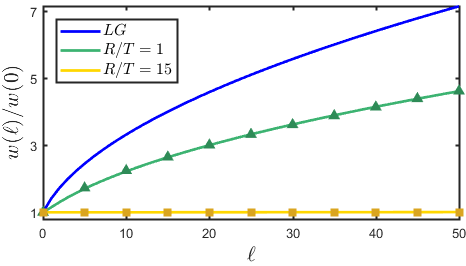}
	\caption{Comparison of the normalised OAM-dependent width between PVs and Laguerre-Gauss beams for different values of $R/T$. For small values of $R/T$, the scaling of the width with $\ell$ approaches that of a Laguerre-Gaussian beam. Lines denote theoretical widths and symbols denote experimental widths.}
	\label{fig:extr}
\end{figure}

In most experiments that generate PVs from Bessel beams, the wavelength of light and the focal length of the Fourier lens are fixed (so that $f$ and $k$ are global constants for all Bessel modes). Consequently, it is worthwhile to see how the width depends on $k_r$ and $w_0$ (the controllable Bessel parameters) independently. This is shown in Figure \ref{fig:krw0}. We experimentally generated PVs with different Bessel beam parameters and compared the widths calculated using Eq.~\ref{eq:comWidth} with Eq.~\ref{eq:PVwidth} for a range of $\ell$ values. We find excellent agreement with theory in all cases. Note that the Bessel beam parameter $k_r$ sets the PV ring radius at $\ell=0$ since $R = k_r f/k$. In a similar way, the Bessel beam parameter $w_0$ sets the PV ring thickness at $\ell=0$ since $T = 2f/k w_0$. Hence, Figs.~\ref{fig:krw0}A and B are equivalent to fixing the PV ring parameters $R$ and $T$, respectively. Again, note that both of these change as $\ell$ changes; the ring thickness decreases for increasing $|\ell|$ whilst the ring radius increases, although the relative change in thickness is usually smaller than the corresponding change in radius.  

From the asymptotic form of the width as given in Eq.~\ref{eq:asympt}, one may be tempted to associate $T$ and $R$ (or equivalently $1/w_0$ and $k_r$) as a psuedo-gradient and psuedo-intercept, respectively, of the OAM-dependent width $w^2(\ell)$. This can perhaps qualitatively explain why each of the curves in Fig.~\ref{fig:krw0}A have different slopes and the same intercept (since $w_0$ varies and $k_r$ is fixed) and why each of the curves in Fig.~\ref{fig:krw0}B have different intercepts but similar slopes (since $k_r$ varies and $w_0$ is fixed). One may then conclude (erroneously) that the Gaussian width $w_0$ of the Bessel beam has the most significant impact on the overall scaling of the PV. However, it's clear from Eq.~\ref{eq:PVwidth} that it is the combined parameter $R/T$ (or $k_r w_0$) which determines the overall scaling of the PV's width with $\ell$. One can verify that if they were to instead plot the normalised width $w(\ell)/w(0)$ in Fig.~\ref{fig:krw0}A and B instead of the absolute width, then the corresponding curves would be identical. This is because corresponding curves in Fig.~\ref{fig:krw0}A and B have the same $R/T$ value. Thus, the value of $R/T$ should be seen as \textit{the} indicator of the degree of ``perfectness'' of quasi-PVs.

As such, care should be taken to ensure that this product is sufficiently large or else the advantages for using PVs are lost. To be more precise, as $R/T \rightarrow 0$, we have that,
\begin{equation}
    w^2(\ell) \approx T^2(\ell+1) + R^2 \,.
\end{equation}
The global scaling is the same as when $R/T \gg 1$, however, since $T$ is now large compared to $R$, the change in width can be significant. This is highlighted in Fig.~\ref{fig:extr} which compares the scaling of LGs with two different PVs: one where $R/T = 15$ and one where $R/T = 1$. This serves to show that if the beam parameters are chosen poorly, the generated PV is far from ``perfect".

\begin{figure}[t] 
	\centering
	\includegraphics[width=\linewidth]{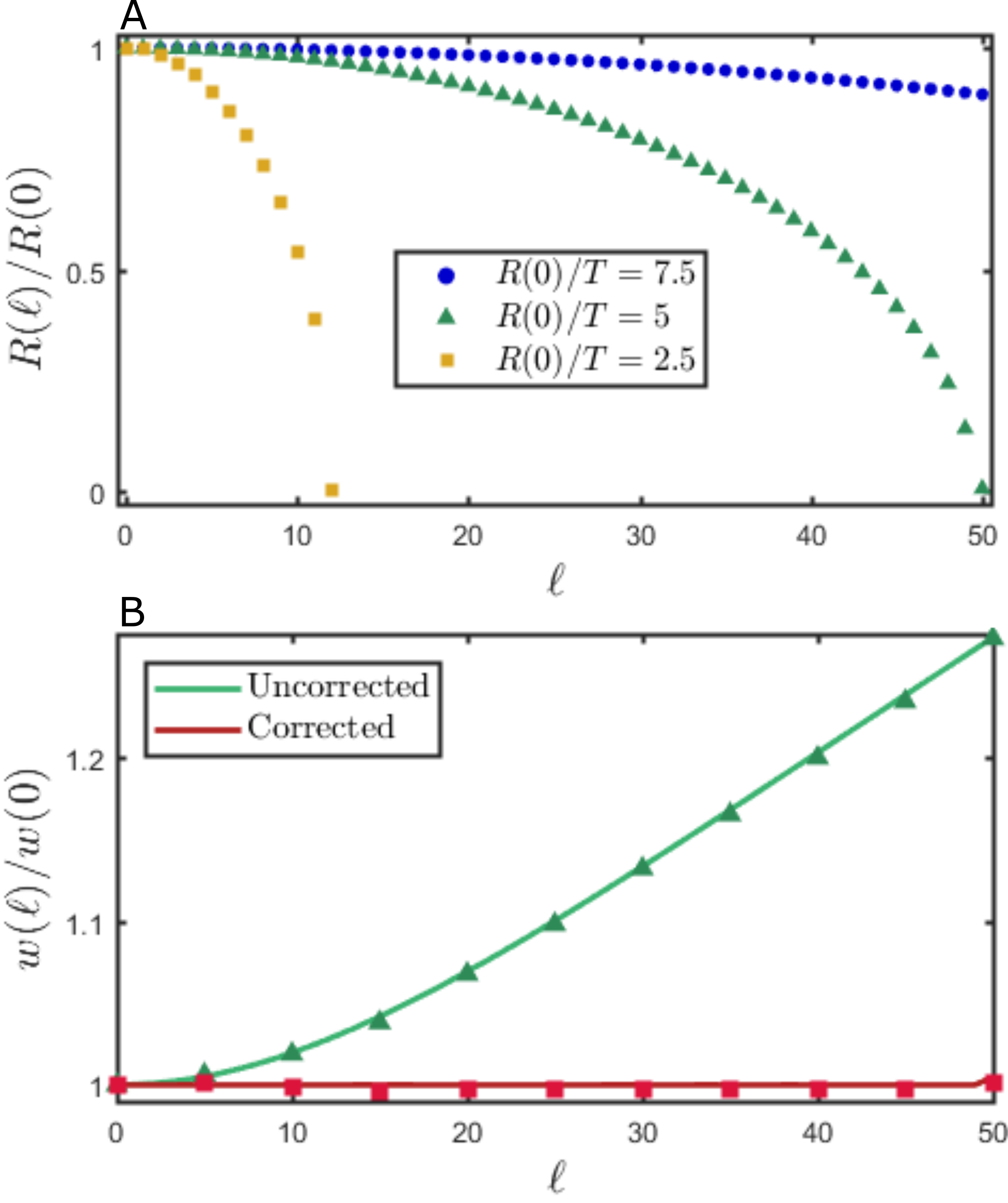}
	\caption{An example of the process of correcting for the width increase by adjusting $R$ individually for each $\ell$. This is equivalent to adjusting the axicon parameter that generates the corresponding BG mode. In A, three examples of the numerical inversion of Eq.~\ref{eq:inversion} are given for different initial radii. In B, we show the uncorrected and corrected experimental widths for the case of $R(0)/T = 5$.}
	\label{fig:constantW}
\end{figure}

Now that the OAM-dependent width of PVs (Eq.~\ref{eq:PVwidth}) is precisely known, we can use this knowledge to compensate for the changing width by adjusting the beam parameters appropriately, thus fixing the PV's width: a sought-after procedure performed in optical trapping experiments which (until now) has been imprecise. Often in such experiments, PVs are generated from BGs and the increasing width is compensated for by digitally tweaking the axicon parameter $\alpha$, which is related to the PV ring radius through the radial wavenumber by $k_r = \alpha(n-1)k$, where $n$ is the axicon's refractive index. Hence, this process effectively compensates for the increasing PV ring radius. Ideally, the the ring thickness should also be adjusted (since we now know that this also changes with $\ell$) but it is much more difficult to dynamically adjust the Gaussian waist in such experiments. Hence, we will assume that the change in the ring thickness is small enough to neglect ($T(\ell)=T(0)=T$). We thus need to find the set of ring radii $R(\ell)$ (or equivalently $\alpha(\ell)$ up to some constant) that will enable $w^2(\ell) = w^2(0)$. This can be done by numerically inverting,
\begin{equation}\label{eq:inversion}
    x^2 \left( \frac{I_{1}\left(x^2 \right)}{I_0\left(x^2 \right)} - \frac{I_{\ell+1}\left(x^2 \right)}{I_\ell\left(x^2 \right)}   \right)  = \ell \,,
\end{equation}
for $x = R(\ell)/T$. Three instances of the numerical inversion are shown in Fig.~\ref{fig:constantW}A and an example of the experimentally corrected PV width for $R(0)/T = 5$ is shown in Fig.~\ref{fig:constantW}B. We see from Fig.~\ref{fig:constantW}A that there is an inevitable cutoff point $\ell_c$ at which point $R(\ell_c) = 0$; the PV ring radius cannot be made any smaller to accommodate the increasing width. We observe that the smaller the initial ring radius $R(0)$ the smaller the cutoff point. As we're about to discuss, a physical explanation for this may be that the OAM density limit has been reached.

It is already known that topological charge densities cannot exist over arbitrarily small areas; in fact, it was derived that the optical vortex density limit within a disk of circumference $2\pi R$ is given by \cite{Roux2003},
\begin{equation} \label{eq:OAMlimit}
    \frac{|\ell|}{R} \leq k \, \text{NA} \,,
\end{equation}
where NA is the numerical aperture of the optical system. This can be interpreted as defining the OAM density limit for a propagating field with a helical phase of the form $\exp(i\ell\phi)$. As the helical phase oscillates faster, higher spatial frequencies are required to maintain this propagating field. Nature, in turn, excites evanescent waves in the region where the OAM density limit is exceeded and so the transverse amplitude within this region decays to zero over a length on the scale of the wavelength. An equivalent interpretation is that for a given $\ell$, Eq.~\ref{eq:OAMlimit} sets the radius $R_\ell$ of the vortex core (the region of the characteristic intensity null in vortex beams), 
\begin{equation} 
R_\ell = \frac{|\ell|}{k\, \text{NA}} \,.
\end{equation}
One can verify that this limit is consistent (to an order of magnitude) with the cutoff radius $R(\ell_c)$. Further, the above shows that the size of the area where evanescent waves are excited is proportional to $|\ell|$. This suggests that a truly OAM-independent beam is unattainable: either the vortex core will eventually engulf the beam or the beam must compensate by growing larger than the vortex core.

In conclusion, we derived an explicit expression for the second moment width of experimentally realisable PVs. The experimental and theoretical results given indicate that the parameter $R/T$ (or $k_r w_0$) primarily dictates the degree of ``perfectness" of quasi-PVs. However, it turns out that even when the ideal PV is best approximated, that is when $R/T \gg 1$, the width will scale in the same way as conventional vortex modes: proportional to $\sqrt{\ell}$. We argued that this is consistent with the already established notion of an OAM density limit, from which it follows that a truly OAM-independent beam is seemingly unattainable. 

\bibliography{mypaperdatabase}



\end{document}